\documentclass[%
reprint,
superscriptaddress,
amsmath,amssymb,
aps,pre,
floatfix,
]{revtex4-2}

\usepackage{graphicx}
\usepackage{dcolumn}
\usepackage{bm}
\usepackage{gensymb}
\usepackage{color}

\usepackage[english]{babel}
\usepackage{siunitx}
\usepackage{physics}
\usepackage[version=4]{mhchem}

\begin{document}

\title{Structural and topological changes across the liquid-liquid transition in water}

\author{Riccardo Foffi }
\affiliation{Department of Physics, Sapienza Universit\`a di Roma, Piazzale Aldo Moro, 2, 00185 Rome, Italy}
\author{John Russo}
\affiliation{Department of Physics, Sapienza Universit\`a di Roma, Piazzale Aldo Moro, 2, 00185 Rome, Italy}
\author{Francesco Sciortino}
\affiliation{Department of Physics, Sapienza Universit\`a di Roma, Piazzale Aldo Moro, 2, 00185 Rome, Italy}

\date{\today}

\begin{abstract}
It has recently been shown 
that the TIP4P/Ice model of water can be studied numerically in metastable equilibrium at and below its liquid-liquid critical temperature.
We report here  simulations along a subcritical isotherm, for which two liquid states with the same pressure and
temperature, but different density, can be equilibrated.
This allows for a clear visualisation of the structural changes taking place across the transition. 
We specifically focus on how the topological properties of the \ce{H}-bond network change across the liquid-liquid transition.
Our results demonstrate that the structure of the high-density liquid, characterised by the existence of interstitial molecules and commonly explained in terms of the collapse of the second neighbour shell, actually originates from
the folding back of long rings, bringing pairs of  molecules 
separated by several hydrogen-bonds close by in space.
\end{abstract}

\pacs{Valid PACS appear here}

\maketitle

The experimental evidence supporting the possibility of a liquid-liquid critical point (LLCP) in supercooled
water is constantly growing~\cite{amann-winkel13-pnas,azouzi2013coherent,perakis2017diffusive,kim2017maxima,stern2019evidence,kim2020experimental}. From the 1992  seminal study by \citet{poole92-nature}
several models~\cite{yamada2002interplay, li2013liquid,ni2016evidence,handle2018potential}, including recently a quantum-based machine-learning  potential~\cite{gartner2020signatures},   have been scrutinized to 
support  the idea that water, despite being a pure molecular liquid, in deep supercooled states  can exist into two different liquid forms.  
Until recently and despite a significant collective effort,
definitive numerical evidence based on the analysis of the density fluctuations close to the 
liquid-liquid critical point and finite size scaling was available only for the  ST2 model~\cite{liu12-jcp,sciortino2011study,kesselring2012nanoscale,palmer14-nature},
a model known to overemphasize tetrahedrality in the local structure. 
Only last year, a numerical tour-de-force investigation~\cite{debenedetti2020second} has finally provided conclusive evidence that two of the most realistic classical models for water~\cite{vega2009ice} 
(TIP4P/2005~\cite{abascal05-jcp} and TIP4P/Ice~\cite{abascal2005potential}) 
do show a clear liquid-liquid critical point in deep supercooled states.  This study, together with the 
previously cited recent outstanding experiments~\cite{amann-winkel13-pnas,azouzi2013coherent,perakis2017diffusive,kim2017maxima,stern2019evidence,kim2020experimental}, strongly supports the possibility that not only the models, but also real water exists in two distinct liquid forms.  Recent reviews~\cite{tanaka2020liquid,salzmann2019advances, handle2017supercooled,gallo16-crev,amann2016colloquium} provide a detailed description of the interesting behavior of supercooled and glassy water.  

The availability of a realistic model that can be numerically studied even below the liquid-liquid critical point, although with a significant computational effort, offers the unprecedented  possibility  to identify the structural units driving the first-order transition, thus  deciphering the microscopic mechanisms at the origin of the  LLCP in water. 

In this manuscript 
we extend previous simulations and investigate novel state points, covering the range from
\( P=\SI{1}{\bar} \) to \( P=\SI{4000}{\bar} \) at \( T=\SI{188}{\K} \) for a system of \( N=1000 \) molecules interacting via the TIP4P/Ice potential.  In this model, the liquid-liquid critical point is found at $T_c=188.6$ K, $\rho_c=1.015$ g/cm$^{3}$, $P_c=1725$ bar ~\cite{debenedetti2020second}.
The possibility to compare the low and high density liquid at the same $P$ and $T$ in metastable equilibrium offers a vivid representation of the structural changes taking place at the transition.  We find that both low and high density liquids are essentially fully bonded, meaning that all
hydrogens ($>99.9$ \%) are engaged in hydrogen bonds with close-by oxygen atoms. The fraction of molecules with four bonds, in the classic two-donors two-acceptors geometry,  is larger than 95\%. Increasing the density across the transition, the tetrahedral geometry is distorted, opening the possibility of 
novel HB patterns, as highlighted by the changes in the  statistics of ring-loops and their spatial geometry.  We discover that the presence of interstitial molecules at relative distances of $0.35$ nm, previously assumed to arise from the collapse of the second shell, originates from the folding back of long rings, bringing molecules separated by paths of several hydrogen bonds at close distances.
Thus pairs of molecules
close by in space ( $\approx 0.35$ nm) are seeds of two distinct networks, facing each other, which merge together after three or four bonds due to the disorder of the network. Introducing a combined structural and topological analysis we show that unambiguous differences in the HB network properties exist between the two liquids.

\section{Numerical methods}
MD simulations were performed in supercooled conditions on the TIP4P/Ice water model~\cite{abascal2005potential} using GROMACS 5.1.4~\cite{abraham2015gromacs}, in the NPT ensemble.
All simulations were run with a leap-frog integrator with a timestep of 
2 fs,  temperature coupling was controlled via the Nos\'e-Hoover thermostat with a characteristic time of \SI{8}{\pico\s}
and pressure coupling via an isotropic Parrinello-Rahman barostat with a characteristic timescale of \SI{18}{\pico\s}.
In all cases a cubic simulation box with periodic boundary conditions was adopted and each simulation was initialized from a different initial configuration.
Molecular constraints were implemented with the LINCS algorithm at sixth order.
Van der Waals interactions have been evaluated with a cutoff distance of $0.9$ nm, and this same cutoff distance was adopted for the real-space cutoff of electrostatic interactions evaluated by the particle-mesh Ewald method at fourth order.
We explored the \( T = \SI{188}{\K} \) isotherm in a system of \( N = 1000 \) molecules, with pressures ranging from \SI{1}{\bar}
to \SI{4000}{\bar}.
All the structural analysis to be presented in the following sections has been performed on the  inherent structures (IS)~\cite{stillinger2015energy,sciortino05-jsm}. The inherent structure of a system can be operationally defined as the  local energy minimum
which is reached via a (constant volume) steepest descent path starting from an equilibrium configuration.  In this process, all vibrational degrees of freedom are brought to their ground state. To numerically evaluate the IS
we used the double-precision version of the steepest descent algorithm in GROMACS with a force tolerance of \( 1 \) J/mol/nm and a maximum step size of \SI{5e-4}{\nano\m}.

\section{Results: Structural and Thermodynamic quantities}

\subsection{Thermodynamic quantities}
\begin{figure}[t]
   \centering
   \includegraphics[width=\columnwidth]{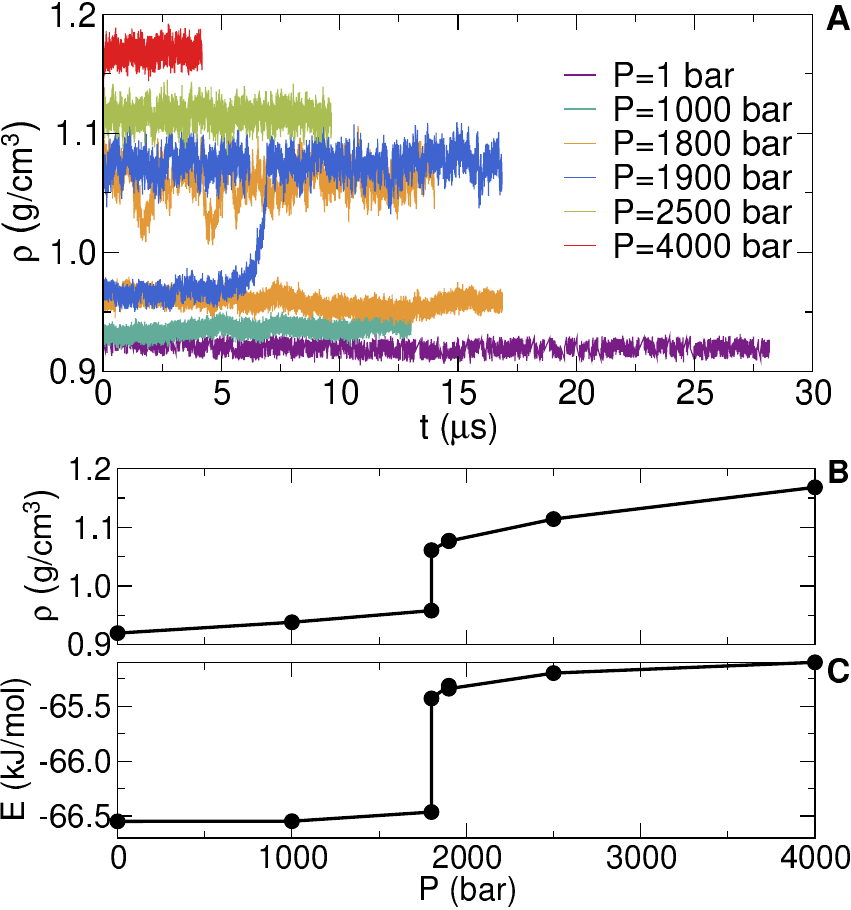}
   \caption{Thermodynamic behaviour along the \( T = \SI{188}{\K} \) isotherm.
   (\textbf A) Fluctuations in the density during the numerical simulations.
   (\textbf B) equation of state and (\textbf C) potential energy per molecule evaluated in the IS. We point out that the jump in energy observed in the IS is identical to that in the real dynamics.}
   \label{fig:Thermodynamics}
\end{figure}

Fig.~\ref{fig:Thermodynamics}A shows the time evolution of the density at \( T = \SI{188}{\K} \)
for different values of $P$.  Close to the liquid-liquid transition (\( P = 1800 \) and \( P = \SI{1900}{\bar} \)) two independent simulations have been performed, starting from opposite sides of the transition line. 
At \( P = \SI{1900}{\bar} \), after \SI{5}{\mu\s}, we observe a transition from low to high density. At \( P = \SI{1800}{\bar} \) no crossing is observed  in the \SI{15}{\mu\s} of
the simulation, providing two trajectories with the same pressure and 
temperature but different densities, as expected close to a coexistence line and, more importantly for us,  two simulations which can be scrutinized to highlight the differences between the low and high density liquid, at the same $P$ and $T$. 

The equation of state at \( T = \SI{188}{\K} \) is shown in Fig.~\ref{fig:Thermodynamics}B. Consistent with  the first-order nature of the transition, the density discontinuously jumps from $\approx 0.95$ to \( \approx \SI{1.05}{\g\per\centi\m^3} \).
A  discontinuous jump of $\approx 1$ kJ/mol is also observed in the pressure dependence of the energy (per molecule) of the inherent structures, reported in Fig.~\ref{fig:Thermodynamics}C.
The condition of  thermodynamic coexistence requires, beside equality of $T$ and $P$ also
the equality of the chemical potential (Gibbs free energy per particle). Hence, from the changes in
$\Delta \rho$ and $\Delta E$ at the transition it is also possible to estimate the change in entropy per molecule $\Delta S$ from
\begin{equation}
\Delta\mu = \Delta E + P\Delta V - T\Delta S = 0.
\end{equation}
From the low density to the high density state at \( T = \SI{188}{\K} \) and \( P = \SI{1800}{\bar} \) we observe $\Delta E \approx\SI{1}{\kilo\J\per\mol}$ and $\Delta V \approx \SI{-2.7}{\AA^3}$ per molecule,
which result in \( \Delta S \approx \SI{3.7}{\J\per\mole\per\K} \).
The high-density liquid is thus more disordered than the low density one, even though it is denser.
Experimental investigations on the corresponding amorphous states LDA and HDA~\cite{whalley1989entropy} report a smaller value,
\( \Delta S \approx \SI{1}{\J\per\mole\per\K} \).
In the two-state model framework~\cite{shi2018common}, entropy differences between pure LDL and HDL for ST2 and TIP5P water have been estimated, respectively, as \( \approx\SI{2.5}{\J\per\mole\per\K} \) and \( \approx\SI{3}{\J\per\mole\per\K} \) on real configurations.
The slope of the coexistence curve at this state point is
\( \eval{\dv*{P}{T}}_\text{coex} = \Delta S/\Delta V \approx \SI{-23}{\bar\per\K} \).

\subsection{Hydrogen Bond}
\begin{figure}[t]
   \centering
   \includegraphics[width=\columnwidth]{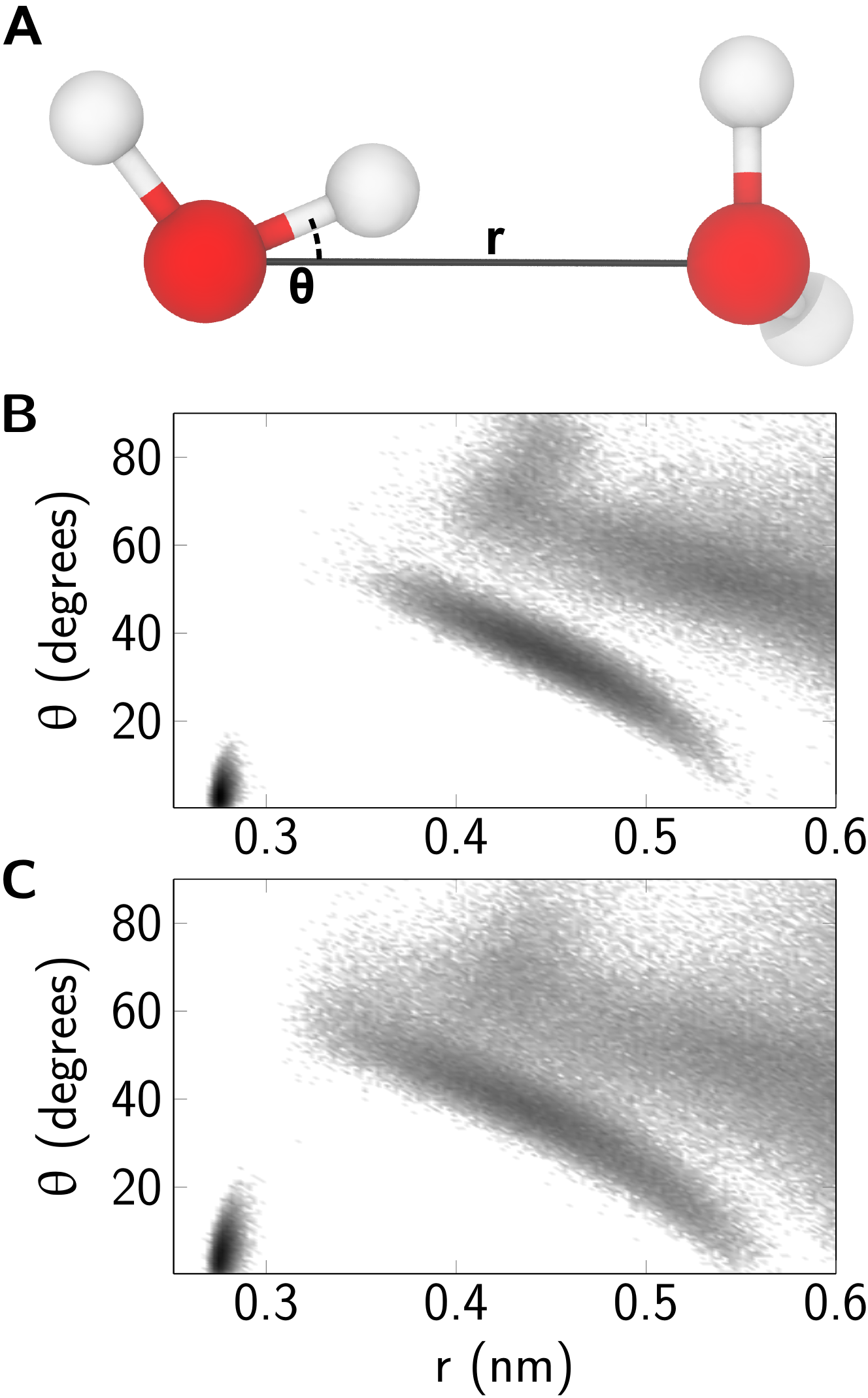}
   \caption{
   (\textbf A) Cartoon representation of intermolecular \ce{O-O} distance \( r \) and minimum \( \ce{H}\hat{\ce{O}}\ce{O} \) angle \( \theta \) used for HB identification.
   Values of $r$ and $\theta$ have been collected(in the IS) between each pair of water molecules to evaluate the joint orientation-distance distribution $P(r,\theta)$.
   Points were sampled from NPT simulations at \( T = 188 \) K and \( P = 1800 \) bar, in (\textbf B) low and (\textbf C) high density
   states. Density $P(r,\theta)$ is represented in log scale.}
   \label{fig:HBonds}
\end{figure}

The ability to provide a proper definition of hydrogen bond(HB) between two molecules is crucial for any analysis of
the hydrogen bond network in water.
Usually, librational and vibrational distortions complicate the HB definition, but their effect can be subtracted by a 
potential energy minimization~\cite{weber1987removing}.
Fig.~\ref{fig:HBonds} shows for each pair of molecules the relative oxygen-oxygen distance \( r \) and the angle \( \theta \), defined as
the smallest among the four \( \ce{H}\hat{\ce{O}}\ce{O} \) angles(the angles between the intramolecular \ce{OH} bond and the intermolecular \ce{OO} line).
The figures show very clearly that at this low \( T \), after minimization, there is a well separated island of points centered
at \( r \approx \SI{0.28}{\nano\m} \) with \( \theta < \SI{30}{\degree} \); this region can be unambiguously associated with hydrogen bonds. In agreement with the geometric criterion of \citet{luzar1996hydrogenbond}, from now on we will define two particles to be \ce{H}-bonded if \( r < \SI{0.35}{\nano\m} \) and \( \theta < \SI{30}{\degree} \) (in the inherent structure configuration).
The more elongated island with \( \theta < \SI{60}{\degree} \) and \( r < \SI{0.55}{\nano\m} \) in the low density state arises from the second shell of neighbors; in the high density state this island expands towards lower values of \( r \) with
increasing \( \theta \), so that the net separation between the shells decreases.
At the same time, it overlaps substantially with the outer one, hinting at a significant spatial mixing between distinct neighbor shells.
More complete analysis of the joint orientation-distance distributions in water
can be found elsewhere~\cite{svishchev1993structure,kusalik1994spatial,kumar2007hydrogen}.

\begin{figure}[t]
	\centering
	\includegraphics[width=\columnwidth]{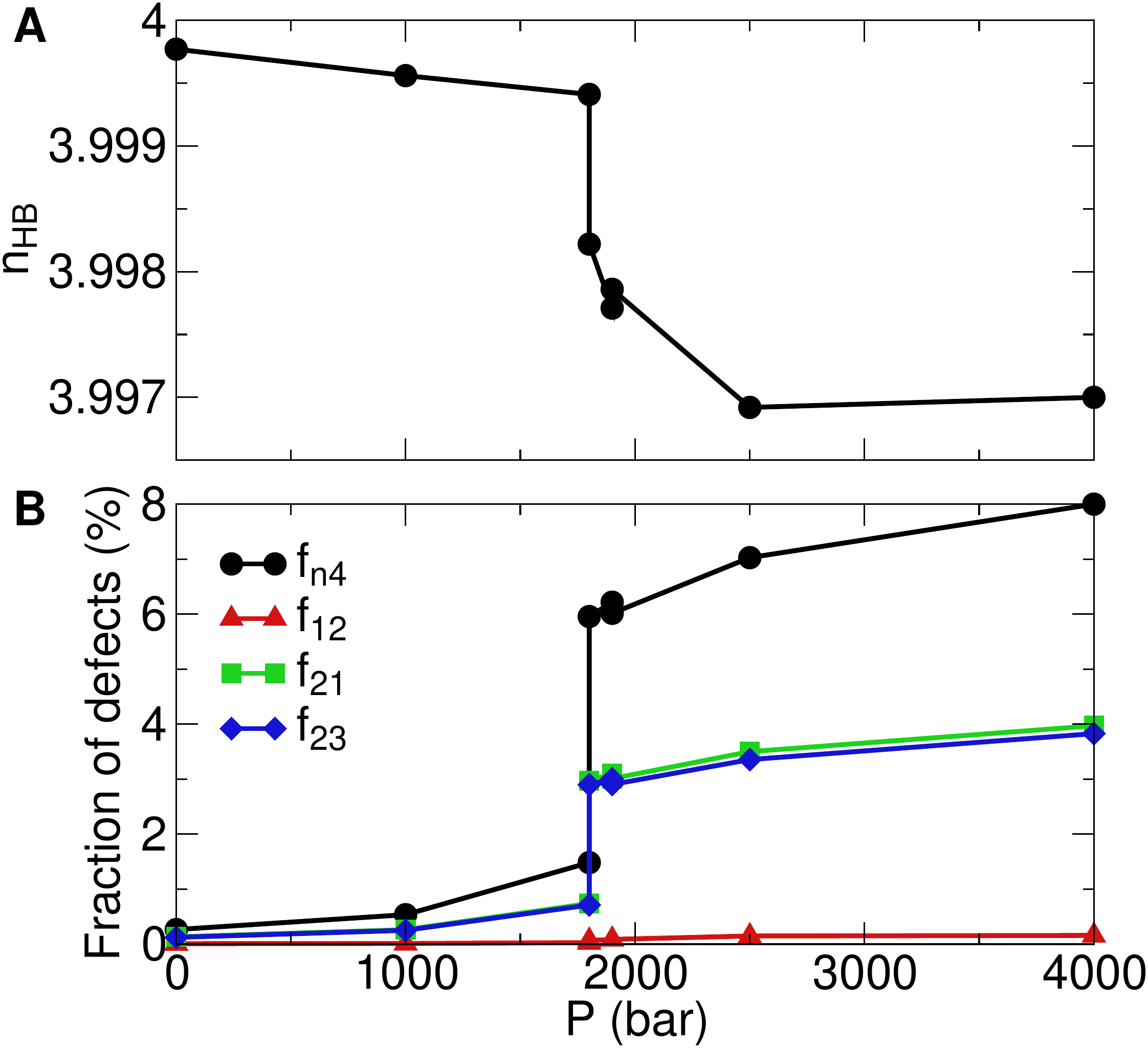}
	\caption{Pressure dependence of
	(\textbf A) the average number of HBs per molecule and
	(\textbf B) the fraction of the different species of coordination defects.
	\( f_{nm} \) is the fraction of \ce{H^nO^m} molecules and \( f_{\text n4} = f_{12}+f_{21}+f_{23} \).}
	\label{fig:HBonds-P}
\end{figure}

The possibility to define the existing HBs unambiguously makes the
analysis of the HB network  quite revealing.  Fig.~\ref{fig:HBonds-P}A shows that at this temperature essentially all
hydrogens are involved in HBs, at all values of $P$.  
The average number of HBs per particle $n_\text{HB}$ is always larger than 3.99. At ambient pressure,
it reaches the value \( n_\text{HB}=3.9997 \), an almost perfect tetrahedral network.
Correspondingly, the fraction of \ce{H} atoms which do not participate in any bond, \( 1-n_\text{HB}/4 \)
(assuming no \ce{H} atom can participate in more than 1 HB),
is \( \approx 7.5\times 10^{-5} \) in LDL and \( \approx 7.5\times 10^{-4} \) in HDL.
The network has reached essentially a fully-bonded state, both in the low and the high density liquid phase.

Despite practically all \ce{H} being involved in one HB, not all molecules show a tetrahedral coordination.  
All  molecules whose bonding pattern deviates from the ideal 2 donated bonds plus 2 accepted bonds can be regarded as network defects.
The idea that water could be described as a defective network of HBs, in which 
bonds break and reform on a  quite fast time scale (ps at ambient temperature) providing
fluidity to the material is not new~\cite{stanley1980interpretation, geiger1979aspects, geiger1982tests, sciortino1989hydrogen}.
The defects have also been associated to molecular mobility~\cite{sciortino1991effect,poole2011dynamical,laage2006molecular,
saito2018crucial}, being the network restructuring
more efficient in the present of such defects acting as catalizers for bond swaps.
In particular, the swap mechanism has been characterized quite precisely in the case of the nearby presence of one over- and one under-coordinated molecule~\cite{laage2006molecular}.
Their structural role and a possible connection with the liquid-liquid phase transition has not been clarified yet.

The defects observed in the explored thermodynamic range are 3- and 5-coordinated molecules. The 3-coordinated molecules coincide with the ``$D$" molecules in Ref.~\cite{montes2020structural}.
Following the notation of \citet{saito2018crucial}, these molecules will be identified, respectively, as
\ce{H^2O^3}, \ce{H^1O^2} and \ce{H^2O^1}, where \ce{H^nO^m} means that the \ce{H} atoms donate $n$ bonds and the \ce{O} atom accepts $m$ bonds.
Fig.~\ref{fig:HBonds-P}B shows the $P$ dependence of the 
fraction of different species  $f_{nm}$ and their sum (the fraction of non four-coordinated molecules $f_{\text n4}$).
In the low density liquid $f_{\text n4}$ is less than 1\%,  all \ce{H} are bonded ($f_{12} \approx 0$) and $f_{21}=f_{23}$.
The low density liquid can thus be represented 
as a fully bonded tetrahedral network, with excitations composed by \ce{H^2O^1 - H^2O^3} pairs, preserving the total number of bonds.   
In the high density liquid, $f_{\text n4}$ is roughly 8\%,  with $f_{21}$ only slightly smaller than $f_{23}$ and a fraction $f_{12} $ smaller than 0.15\% of un-bonded hydrogens.
Hence, even the high density liquid has a large majority (\( >92\% \)) of tetrahedrally-coordinated molecules.
At the transition, the number of non-tetrahedral coordinated molecules jumps from $\approx 2\%$ to  $\approx 6$\%.
We never observe \ce{H^3O^2} molecules, in which the same proton binds to two different oxygens~\cite{giguere1987bifurcated}.
In the study of \citet{saito2018crucial}, which focuses on network defects on a wider thermodynamic range, no mention of
\ce{H^3O^2} has been made;
other works using the ``bifurcated bond" nomenclature~\cite{sciortino1991effect,laage2006molecular} do not distinguish explicitly between \ce{H^2O^3} and \ce{H^3O^2}.

We note on passing that, as we have alluded previously,
\ce{H^1O^2} describes an un-bonded proton, a geometry that sacrifices one HB and whose probability becomes rarer and rarer on cooling.    
It has been suggested~\cite{saito2018crucial} that
the \ce{H^1O^2} defects  play a more central role as a promoter of structural changes, facilitating  rotational and translational motions in water.   Hence, despite vanishingly rare, it could retain some fundamental role in the glassy dynamics of water and explain the higher mobility of the high density liquid compared to the low density one.
The almost perfect matching of $f_{23}$ and $f_{21}$ at this extremely low temperature also suggests that possibly these network defects are created in pairs and, once created, propagate in the network.  
Finally, we also note that in the low density phase at atmospheric pressure,
configurations with no defects (out of 1000 molecules) are sampled during the course of the simulation, providing to our knowledge the first {\it brute force} generation of a perfect random tetrahedral network in water.

\begin{figure}[t]
	\centering
	\includegraphics[width=\columnwidth]{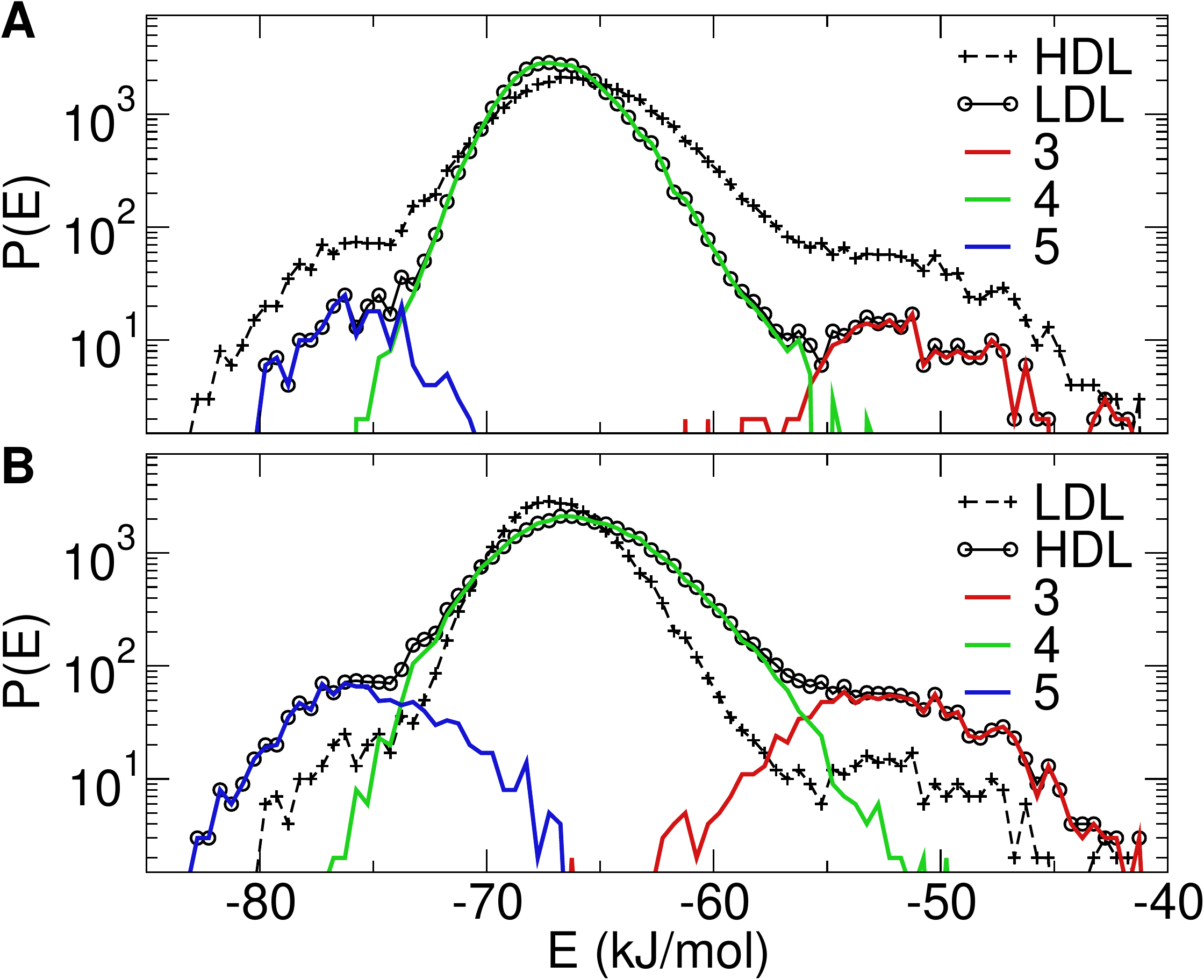}
	\caption{Distribution of potential energy per molecule in the inherent structures at
	\( T = \SI{188}{\K} \) and \( P=\SI{1800}{\bar} \). Each of the two panels shows the contribution from both LDL and HDL for better comparison. In \textbf A, the separation in the contributions from 3-, 4-, and 5-coordinated molecules is referred to the LDL; in \textbf B it is referred to the HDL.}
	\label{fig:EHist}
\end{figure}

Fig.~\ref{fig:EHist} shows the distribution of the potential energy per molecule in the IS at \( P = \SI{1800}{\bar} \) for both the low and the high density liquids, separating the contributions from molecules with different coordination numbers.  It is apparent from the data that five-(three-)coordinated molecules have a lower (higher) energy compared to the four-coordinated,
supporting the view  that the combined creation of a  \ce{H^2O^1} and a \ce{H^2O^3} has a relatively small energetic cost, since it 
it does not require a change in the total number of hydrogen bonds.
We also note that there is a net effect of the density on the tetrahedrally-coordinated molecules(\ce{H^2O^2}): as the density increases, their energy distribution displays an increase in variance towards the high-energy side.
The average energy variation per molecule from LDL to HDL is \( \Delta E \approx \SI{1}{\kilo\J\per\mole} \);
the separate contributions based on coordination number are
\( \Delta E_4 \simeq \SI{0.95}{\kilo\J\per\mole} \),
\( \Delta E_3 \simeq \SI{-0.96}{\kilo\J\per\mole} \),
\( \Delta E_5 \simeq \SI{0.46}{\kilo\J\per\mole} \).
3-coordinated molecules are favored in the HDL state, while 5-coordinated molecules are favored in the LDL state.
Assuming for convenience that defects are created \emph{exactly} in pairs in both states, defining the average energy of a defect
\( (E_3+E_5)/2\equiv E_\text{n4} \), and denoting the HDL and LDL states with superscripts H and L respectively,
\begin{equation*}
	\begin{split}
		\Delta E = E^\text{H} - E^\text{L}
		=& \left( 1-f_\text{n4}^\text{L} \right)\Delta E_4 + f_\text{n4}^\text{L}\Delta E_\text{n4} + \\
		&+ \Delta f_\text{n4} E_\text{n4}^\text{L} - \Delta f_\text{n4} E_4^\text{L} + \\
		&+ \Delta f_\text{n4} \left( \Delta E_\text{n4} - \Delta E_4 \right) = \\
		=&\; \Delta E_\rho + \Delta E_\text{def} + \Delta E_\text{c}.
	\end{split}
\end{equation*}
Three distinct contributions to the energy variation are identified: density increase(and hence network distorsion), creation of defect pairs, and a coupling term. Extracting the relevant quantities from the MD trajectories, one obtains
\( \Delta E_\text{n4} \approx \SI{-0.25}{\kilo\J\per\mole} \),
\( E_4^\text{L} \approx \SI{-66.8}{\kilo\J\per\mole} \),
\( E_\text{n4}^\text{L} \approx \SI{-63.5}{\kilo\J\per\mole} \),
\( f_\text{n4}^\text{L} \approx 1.5\% \),
\( \Delta f_\text{n4} \approx 4.5\% \);
therefore the three contributions are
\( \Delta E_\rho \approx \SI{0.93}{\kilo\J\per\mole} \),
\( \Delta E_\text{def} \approx \SI{0.15}{\kilo\J\per\mole} \) and
\( \Delta E_\text{c} \approx \SI{-0.05}{\kilo\J\per\mole} \).
The two components of the defect contribution are large but mostly cancel each other out since the energy of a 3-5 defect pair is only slightly higher than that of a tetrahedral pair.
The dominant effect of the density   increase is a distortion of the HB network structure; even though defects gain some stability from the network distortion (\( \Delta E_\text{n4}<0 \)), their concentration is too low to counterbalance the energy increase in the distortion of the tetrahedrally-coordinated molecules.

\subsection{Structural quantities}

\begin{figure}[t]
	\centering
    \includegraphics[width=\columnwidth]{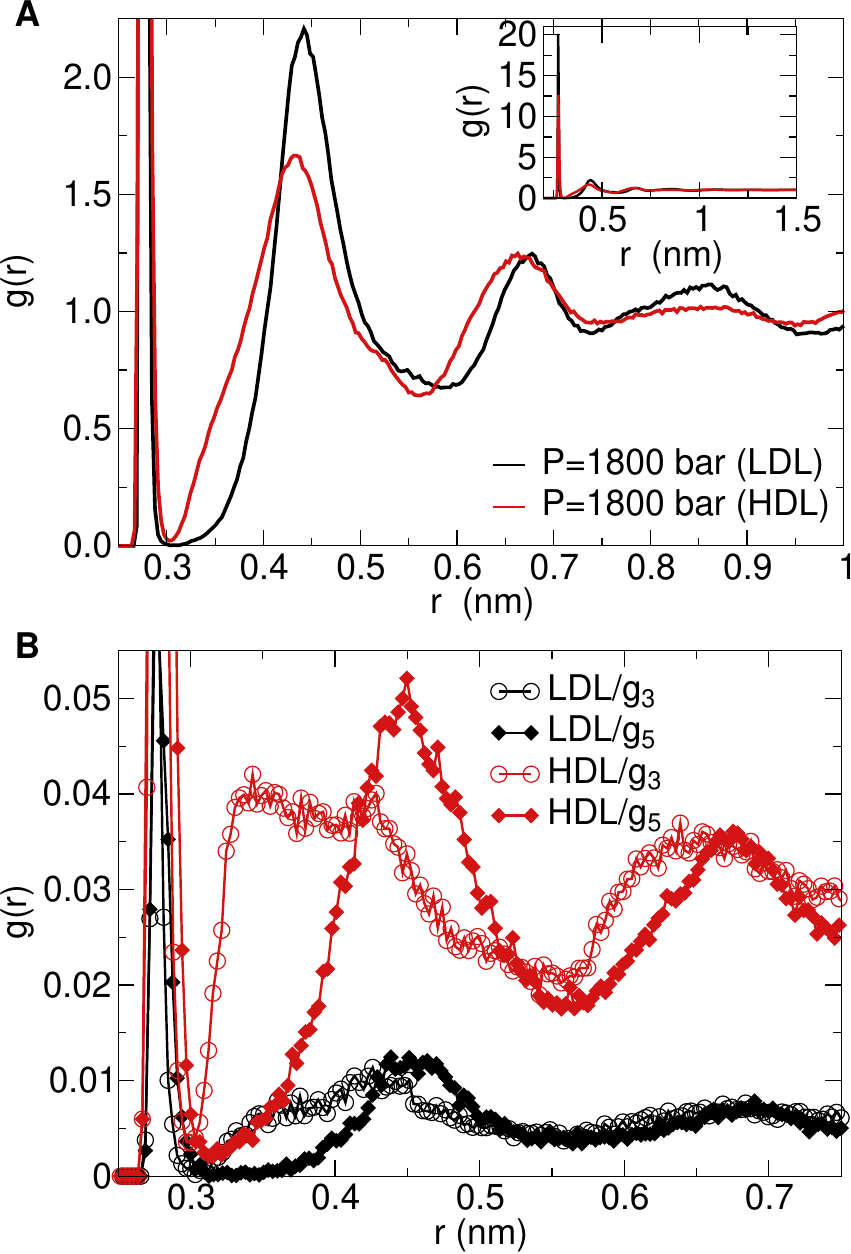}
    \caption{Structural differences between the LDL and HDL state at \( T=\SI{188}{\K} \) and \( P=\SI{1800}{\bar} \).
    (\textbf A) Total oxygen-oxygen radial distribution \( g(r) \) in the two states.
    (\textbf B) Contribution of the defects to \( g(r) \) in the two states.} 
    \label{fig:gr}
\end{figure}

Fig.~\ref{fig:gr}A shows the oxygen-oxygen radial distribution function $g(r)$ (evaluated in the IS) in the low and high density liquid at \( P = \SI{1800}{\bar} \). 
Having subtracted the vibrational and librational components, both states show
\( g(r) \approx 0 \) at \( r \approx \SI{0.31}{\nano\m} \), an indication of  a very well defined first coordination shell.
For \( r > \SI{0.3}{\nano\m} \) clear differences emerge between the two liquids:
the high density liquid displays a significant presence of interstitial molecules between the first and second coordination shell,
while the low density liquid has a well-defined second peak at 
a distance of $\approx \SI{0.44}{\nano\m}$, consistent with a tetrahedral bonding geometry. A large amount of studies, both numerical and experimental, have focused on understanding the characteristic features of distinct structures of liquid water. It is now acknowledged that the main difference between the LDL and the HDL lies in the region between the first two minima of the oxygen-oxygen $g(r)$, i.e. the second shell~\cite{gallo2016water,tanaka2019revealing}; in particular, while the second shell of the low-density structure is consistent with an optimal tetrahedral arrangement, the second shell of the high-density one is observed to `collapse' towards the first\cite{soper2000structures}.

To estimate the effect of network defects on the average structural quantities, we evaluate \( g_3 \) and \( g_5 \), the \( g(r) \) evaluated around a reference molecule with three and five HBs, respectively. Both 
 \( g_3 \) and \( g_5 \) are normalized such that the large $r$ limit coincides with their relative concentration, to highlight their relevance in the total $g(r)$. 
Fig.~\ref{fig:gr}B shows that the local environment of three-coordinated molecules is significantly populated
at \( r \approx \SI{0.35}{\nano\m} \) in both LDL and HDL states;
although it is much weaker, also five-coordinated molecules show some tendency to populate the interstitial region, but only in the HDL.
However  Fig.~\ref{fig:gr}B shows  that due to their small concentration, coordination defects' net contribution to the \( g(r) \) is practically irrelevant: their presence cannot explain, alone, the presence of interstitial molecules in HDL. The vast majority of interstitial molecules must thus be tetrahedrally-coordinated molecules surrounding other tetrahedrally-coordinated molecules.

\section{Results: Rings and Network topology}
\subsection{Particle-rings}
\label{sec:particlering}
The possibility to define unambiguously the hydrogen bonded pairs  at the explored temperatures (Fig.~\ref{fig:HBonds}) offers the possibility to
investigate the topology of the HB network and its changes across the LL transition.
We will consider the HB network as undirected, and our investigation will make use of two basic concepts.
The first is the chemical distance \( D \) between two molecules, defined as the length of the shortest HB path connecting them;
the set of molecules at the same chemical distance \( D \) from a central molecule is said to constitute its \( D \)-th {\it bond-coordination shell}, which is to be distinguished from the notion of coordination shell as identified by the minima of \( g(r) \).
The second concept is that of ring, that we define as a closed path along the HB network, whose total length \( L \) is measured in units of chemical distance.
The analysis of  the ring statistics is not  new  in the structural studies of network-forming materials, and many different non-equivalent definitions of ``ring" (or loop) have been proposed throughout the last 50 years, providing sometime contrasting physical interpretations~\cite{rahman1973hydrogenbond,speedy1987network,formanek2020probing}. 

To provide a characterization of the local bonding environment around a given molecule \( i \), we evaluate, for each pair \( (j,k) \) of first-bonded neighbors of \( i \)
(i.e. at chemical distance one), the shortest closed path, starting and ending in \( i \), that contains both \( j \) and \( k \). This closed path will be called a \emph{particle-ring} of molecule $i$.
This is equivalent to summing the two shortest non-intersecting paths starting at 
\( j \) and ending at \( k \) (one of which is always the path $j-i-k$)
The length of the particle-ring is evaluated in units of chemical distance.
Using this definition, a tetrahedrally-coordinated molecule  has exactly six 
particle-ring lengths associated to it.  Instead,  a three(five)-coordinated molecule has three(ten) particle-ring lengths.

\begin{figure}[t]
    \centering
    \includegraphics[width=\columnwidth]{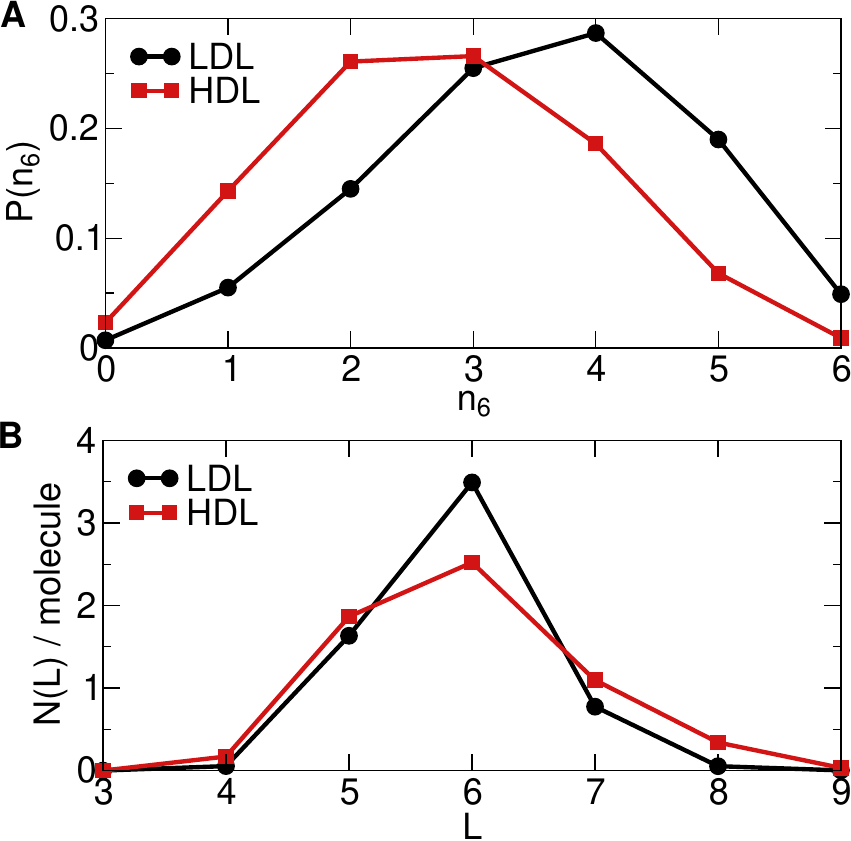}
    \caption{Particle-ring statistics in the LDL and HDL state at \( T = \SI{188}{\K} \) and \( P = \SI{1800}{\bar} \).    (\textbf A) Fraction of tetrahedrally-coordinated molecules with \( n_6 \) hexagonal rings steaming out from their four bonds     (\( 0 \le n_6 \le 6 \)).    (\textbf B) Distribution of particle-ring lengths, normalized to represent the average number of rings of given length per molecule.}
    \label{fig:rings}
\end{figure}

In a crystalline ice I\textsubscript{h} configuration, all six particle-rings emanating from a generic molecule have length six.
The disorder of the liquid phase manifests in the existence of a continuum of local molecular environments, resulting in a distribution of particle-ring lengths. 
Fig.~\ref{fig:rings}A shows the fraction $P(n_6)$ of tetrahedral molecules with a given number \( n_6 \) of hexagonal particle-rings in the two liquid states. 
The HDL has on average $\approx 2.5$ hexagonal particle-rings while the LDL has on average $\approx 3.5$ of them. 

Particularly interesting is the distribution of particle-ring lengths in the low- and high-density liquid at \( P = \SI{1800}{\bar} \), shown in  Fig.~\ref{fig:rings}B. The LDL has a larger population of hexagonal particle-rings and a smaller fraction of five- and seven-membered ones
(with negligible concentrations of \( L=4 \) and \( L=8 \)).
The HDL is instead characterized by a more disordered particle-ring distribution, with a non-negligible presence of particle-rings of length four and eight and a few of length nine. 
Thus, particle-rings of length four and eight could act as possible indicators of typical HDL local environments.

\begin{figure}
    \centering
    \includegraphics[width=\columnwidth]{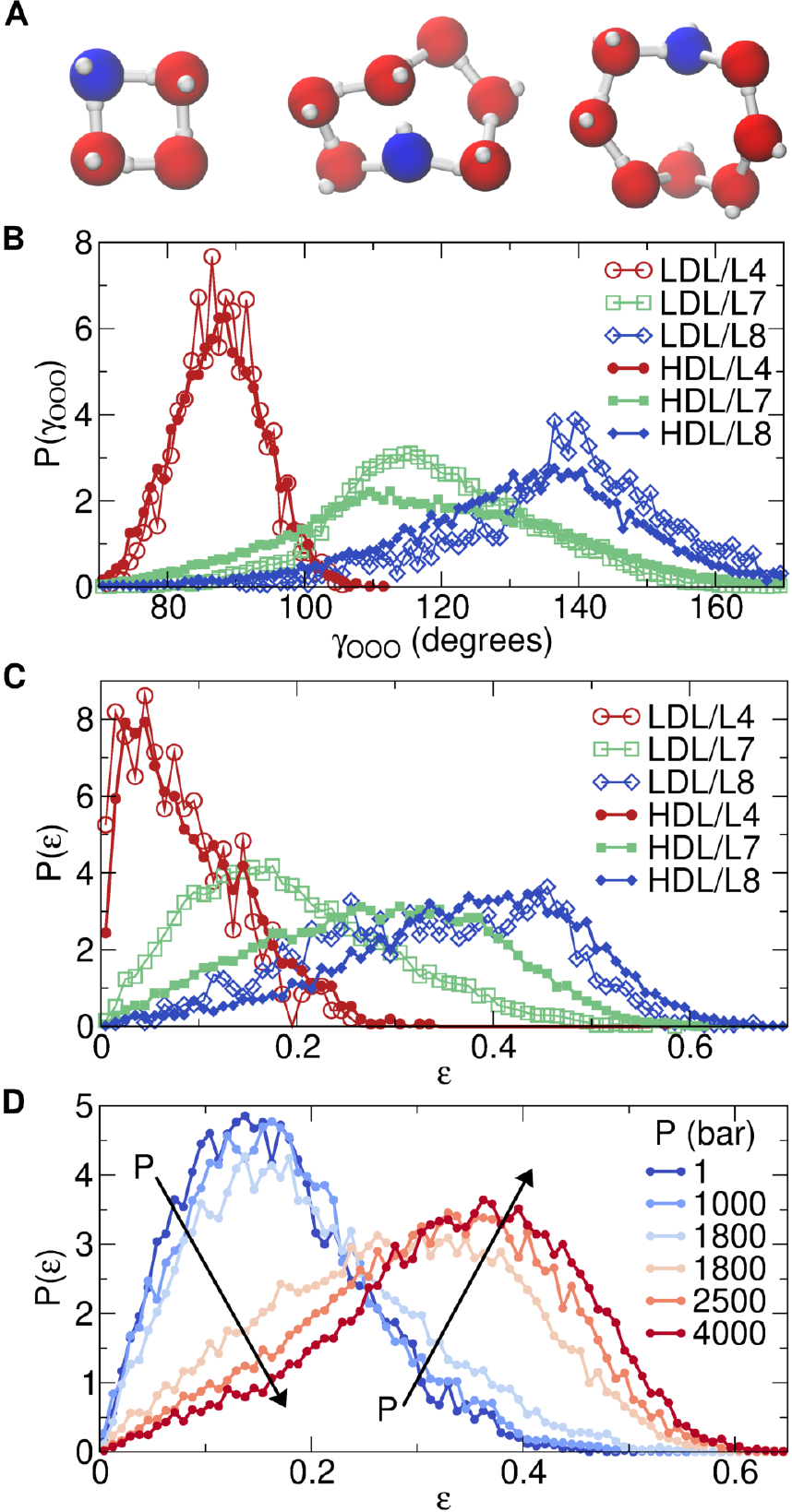}
    \caption{Geometric properties of particle-rings.
    (\textbf A) Cartoon representation of particle-rings of length 4, 7 and 8 with the seed molecule highlighted in blue.
    (\textbf B) Distribution of the \ce{O-O-O} angles of particle-ring seeds and
    (\textbf C) of the elongation factor of particle-rings in the LDL and HDL states at
    $T=\SI{188}{\K}$ and $P=\SI{1800}{\bar}$.
    (\textbf D) $P(\varepsilon)$ in particle-rings of length 7 along the \SI{188}{\K} isotherm shows a clear bimodal nature.
    All curves are normalized so that the underlying areas don't reflect the concentration of each population.}
    \label{fig:ParticleRingShape}
\end{figure}

Representative images of particle-rings of various lengths are shown in Fig.~\ref{fig:ParticleRingShape}A.  To quantify the variability in their morphological properties we evaluate two geometric quantities: the \ce{O-O-O} angle of the seed molecule, $\gamma_\text{OOO}$, and the elongation factor of the ring $\varepsilon = (I_2-I_1)/I_3$ where $I_n$ is the $n$-th eigenvalue of the ring inertia tensor(sorted by increasing magnitude)~\cite{martonak2005evolution}.
$\varepsilon$ has values between zero for a radially symmetric arrangement and one for a linear arrangement. The results are shown in 
Fig.~\ref{fig:ParticleRingShape}B-C, for four-, seven-, and eight-membered particle-rings.
Tetragonal rings are unique in that they have a well-defined shape, for which only two types of arrangements can be identified: one in which each molecule in the ring accepts and donates one HB to its in-ring neighbors, and one in which two molecules act as both donor and acceptor while the other two act only as donor or acceptor respectively. The morphology of these particle-rings is identical in both liquid states, suggesting that
these rings are not affected by the transition.
Heptagonal particle-rings are significantly present in both LDL and HDL(Fig.~\ref{fig:rings}) and show a very broad $\gamma_\text{OOO}$ distribution, hinting at the fact that there is no strict geometric constraint on their formation and they can adapt to various shapes. This is also confirmed by their $P(\varepsilon)$, showing that they can range from ``circular" configurations to folded and elongated ones. 
Octagonal particle-rings have seeds with large characteristic $\gamma_\text{OOO}$ angles, sometimes reaching configurations with almost-collinear triplets of oxygens, although their distribution is broad and rings with tighter angles are also found.
Their distributions does not undergo a drastic change across the transition, but, while they clearly favor elongated configurations, they can still adopt a variety of morphologies, as implied by the long tails of $P(\varepsilon)$.
Seven-membered particle-rings are the ones to undergo the most drastic morphological transformation across the transition; the full pressure dependence of their $P(\varepsilon)$ is drawn in Fig.~\ref{fig:ParticleRingShape}D, showing a distinct bimodal nature. While their concentration increases only slightly with increasing density, most of them deform to take on elongated shapes, so that the HDL is dominated by elongated rings: the average elongation of seven-membered particle-rings goes from $\approx 0.16$ at ambient pressure up to $\approx 0.34$ at \SI{4000}{\bar}.
All the other particle-rings(except tetragonal) show a similar tendency to increase $\epsilon$, but not as clear and pronounced.

\subsection{Merging Structure and Network properties}

\begin{figure}[t]
	\centering
	\includegraphics[width=\columnwidth]{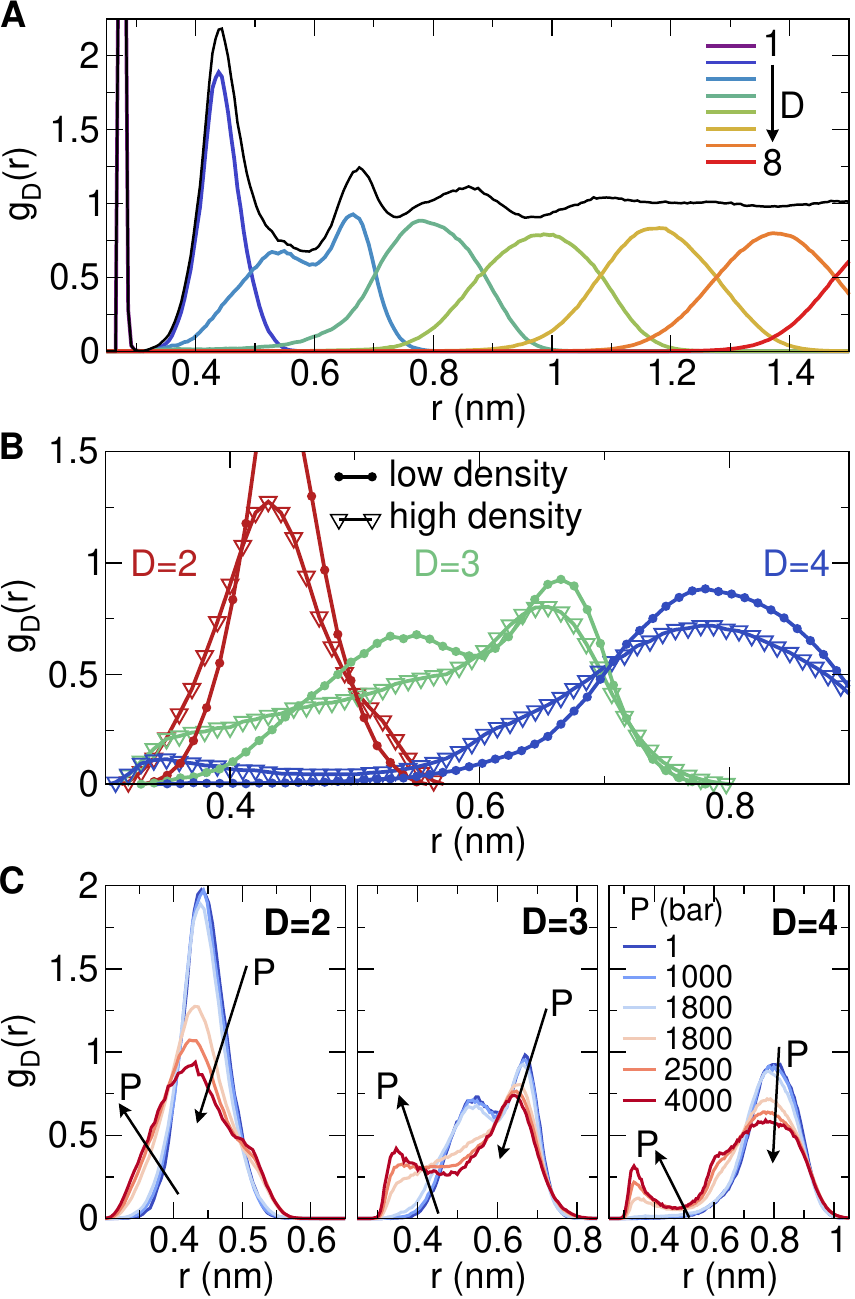}
    \caption{ 
    (\textbf A) \( g(r) \) of water(black) splitted in the contributions arising from each distinct
    bond-coordination shell from \( D=1 \) to \( D=8 \) in LDL at \( P=\SI{1800}{\bar} \).
    (\textbf B) Closeup on the \( g(r) \) contributions from second to fourth bond-coordination shells.
    Curves have been evaluated from simulations at \( T=\SI{188}{\K} \) and \( P=\SI{1800}{\bar} \),
    in the low and high density state. Colors are matched per bond-coordination shell.
    (\textbf C) Pressure dependence of the \( g(r) \) contributions from second, third and fourth bond-coordination shell along the \SI{188}{\K} isotherm.
    }
    \label{fig:RDF-ChemD}
\end{figure}

\begin{figure*}[t]
    \centering
    \includegraphics[width=.95\textwidth]{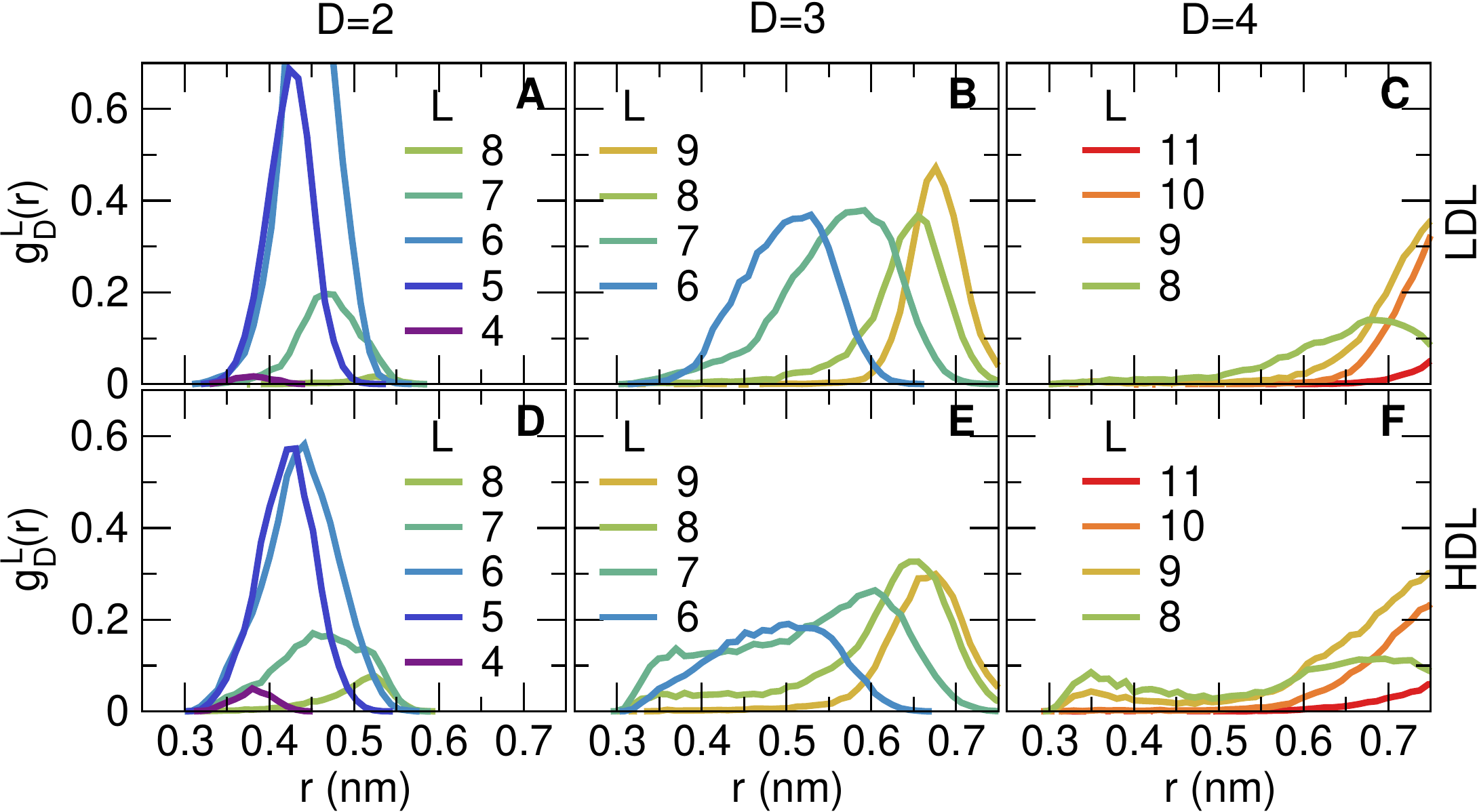}
	\caption{Contributions to the oxygen-oxygen $g(r)$  of water separated by chemical distance \( D \) and ring length \( L \) in the low-density (\textbf A--\textbf C) and high-density (\textbf D--\textbf F) states at \( P=\SI{1800}{\bar} \) and \(T = \SI{188}{\K}\).
	Of particular interest are the clear peaks for $D=4$ in the interstitial region of the high density liquid (\textbf F).}
    \label{fig:RDF-Rings}
\end{figure*}

It is well known  that the second coordination shell, i.e. the spherical shell formed by neighbours at a distance between the first and second minima of the radial distribution function from the central atom is crucial for a structural description of liquid water~\cite{gallo2016water,tanaka2019revealing,soper2000structures}.
In the case of network-forming liquids, we find particularly informative to distinguish between {\it shells} in real space distance as commonly defined by the minima of the \( g(r) \)~\cite{pathak2019intermediate}, and {\it bond-coordination shells}, as indicated by the chemical distance. This network-inspired separation of the different 
contributions of the radial distribution function provides a novel unexpected result: the main contribution to the collapse of the second shell observed in real space, which is associated to the LL transition, actually originates from the ``folding" of rings longer than six,
which brings pairs of topologically distant molecules, in particular those at chemical distances three and four, at short distances in space.
 To quantify this observation,   we start by evaluating the contribution of each distinct bond-coordination shell to 
\( g(r) \), so that \( g(r) = \sum_D g_D(r) \) where
\( g_D(r) \) is the radial distribution function evaluated among only pairs of molecules at chemical distance \( D \).
A sample result of this procedure, applied in the low density state at \SI{1800}{\bar}, is shown in Fig.~\ref{fig:RDF-ChemD}A.
One should notice that {\it peaks in the \( g(r) \) do not directly mirror peaks in the underlying \( g_D \) contributions}.
Related to the interstitial molecule phenomenon ($r \approx \SI{0.35}{\nano\m}$) are the distributions for chemical distances \( 2\le D\le 4 \), shown in Fig.~\ref{fig:RDF-ChemD}B in both LDL and HDL at the same $P$.
The increase in density  couples with the broadening of all coordination shells and, interestingly, to the significant growth of a population in the third and fourth bond-coordination shell at distances corresponding to the second spatial shell.
These changes become even more clear if the full pressure dependence of \( g_D \) along the
\( T=\SI{188}{\K} \) isotherm is examined, as shown in Fig.~\ref{fig:RDF-ChemD}C, for each of these bond-coordination shells.
Upon increasing \( P \) the second bond-coordination shell significantly broadens,
populating  both distances smaller and larger than the typical tetrahedral geometry, with a net 
shifts towards lower separation values.
But the most striking result is the growth in the HDL of well-defined peaks in the interstitial region (\( \approx\SI{0.35}{\nano\meter} \)) arising from molecules at chemical distances
\( D=3 \) and \( D=4 \).
These peaks are thus an intrinsic feature of the high density liquid and 
a clear evidence of a structural change in the network topology across the LDL-HDL transition.

\begin{figure*}
    \centering
    \includegraphics[width=.98\textwidth]{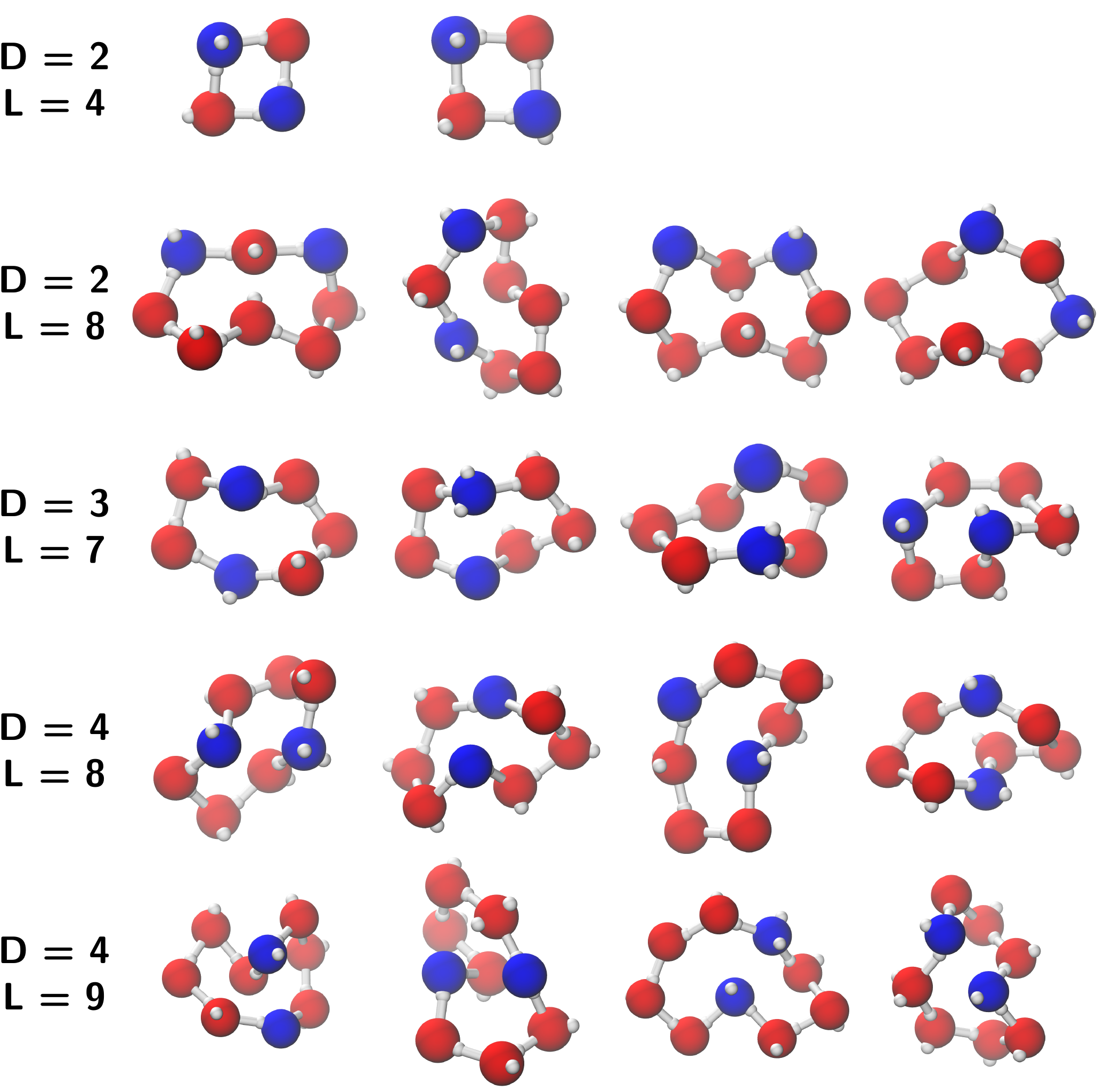}
    \caption{Representative images of different types of rings highlighting some of the structural features of the high-density liquid. The blue molecules are the ones between which the distance $r$ is calculated. The D2L4 structures are more square-like if all four participating molecules donate at most one HB, more rhomboid if one of the molecules donates two HBs, imposing a tetrahedral angle. The D2L8 rings have been chosen such that  $r > \SI{0.5}{\nano\m}$, to highlight the highly distorted, quite collinear, elongated bonds.
    All the other examples (D3L7, D4L8, D4L9) have $r < \SI{0.37}{\nano\m}$, so that each blue molecule is an interstitial molecule for the other one.}
    \label{fig:RingsView}
\end{figure*}

A closer look at the details of this topological transformation is obtained by further separating the radial distributions $g_D$ in the contributions arising from different ring lengths.  Here for each pair of molecules at chemical distance \( D \) we evaluate the length of the two shortest non-intersecting paths connecting the two molecules and define the ring length \( L \) as the sum of these two lengths. 
This definition of \emph{ring} is not to be confused with that of particle-ring given previously, since no central particle is involved (although rings associated to particles with $D=2$ do coincide with particle rings for any ring length $L$).
We then separate each \( g_D(r) \) into the contributions from each ring population
\( g_D(r) = \sum_L g_D^L(r) \).
By definition, the minimum ring length between two molecules at chemical distance \( D \) will satisfy \( L \ge 2D \).
To clarify more the methodology, we show in  Fig.~\ref{fig:RingsView}  a collection of rings characteristic of the HDL state with their classification in term of chemical distance $D$ and minimum ring length $L$.  The two blue molecules are the ones
for which the distance in real space $r$ is calculated.

The resulting radial distribution function $g_D^L$, evaluated on the LDL and HDL states at \SI{188}{\K} and \SI{1800}{\bar},
are shown in Fig.~\ref{fig:RDF-Rings}.
Our discussion will start from the second bond-coordination shell(panels A and D). As we move from LDL to HDL we observe, along with a drastic decrease in the population of hexagonal rings(expected, and already evident from Fig.~\ref{fig:rings}), a broadening in the contribution of heptagonal rings, a symptom of the increased disorder.
What is probably most interesting is, however, the presence of tetragonal and octagonal rings in the HDL, which are nearly absent in the LDL.
Tetragonal rings are distorted ``square" configurations whose diagonals contribute to the peak around
\( \approx\SI{0.38}{\nano\m} \), implying that the \ce{O-O-O} angles between each triplet of adjacent molecules lie around an average of \( \approx \SI{87}{\degree} \).
Pairs at \( D=2 \) in octagonal rings show instead a clear peak at distances
\( \approx \SI{0.52}{\nano\m} \), indicating almost-collinear triplets of H-bonded molecules, a highly distorted arrangement.

In the third bond-coordination shell (panels B and E) the population of hexagonal rings diminishes but develops a fatter low-\( r \) tail, and a major distortion in both heptagonal and octagonal structures is observed, populating the interstitial region; longer rings also show extending tails but their contribution in the region of the second spatial shell is negligible.
  
The situation in the fourth bond-coordination shell (panels C and F)
is particularly revealing. 
Both octagonal and enneagonal rings grow a secondary peak centered around
\( r\approx\SI{0.35}{\nano\m} \) in  HDL. Larger rings develop long tails but contribute only negligibly to the second spatial shell.
The peaks from eight- and nine-membered rings provide a contribution in the interstitial region that is completely missing in the LDL phase; this can only happen if these rings fold back on themselves, so that topologically distant pairs of molecules on the opposite sides of the ring are brought close to each other.

\begin{figure*}[t]
    \centering
    \includegraphics[width=\textwidth]{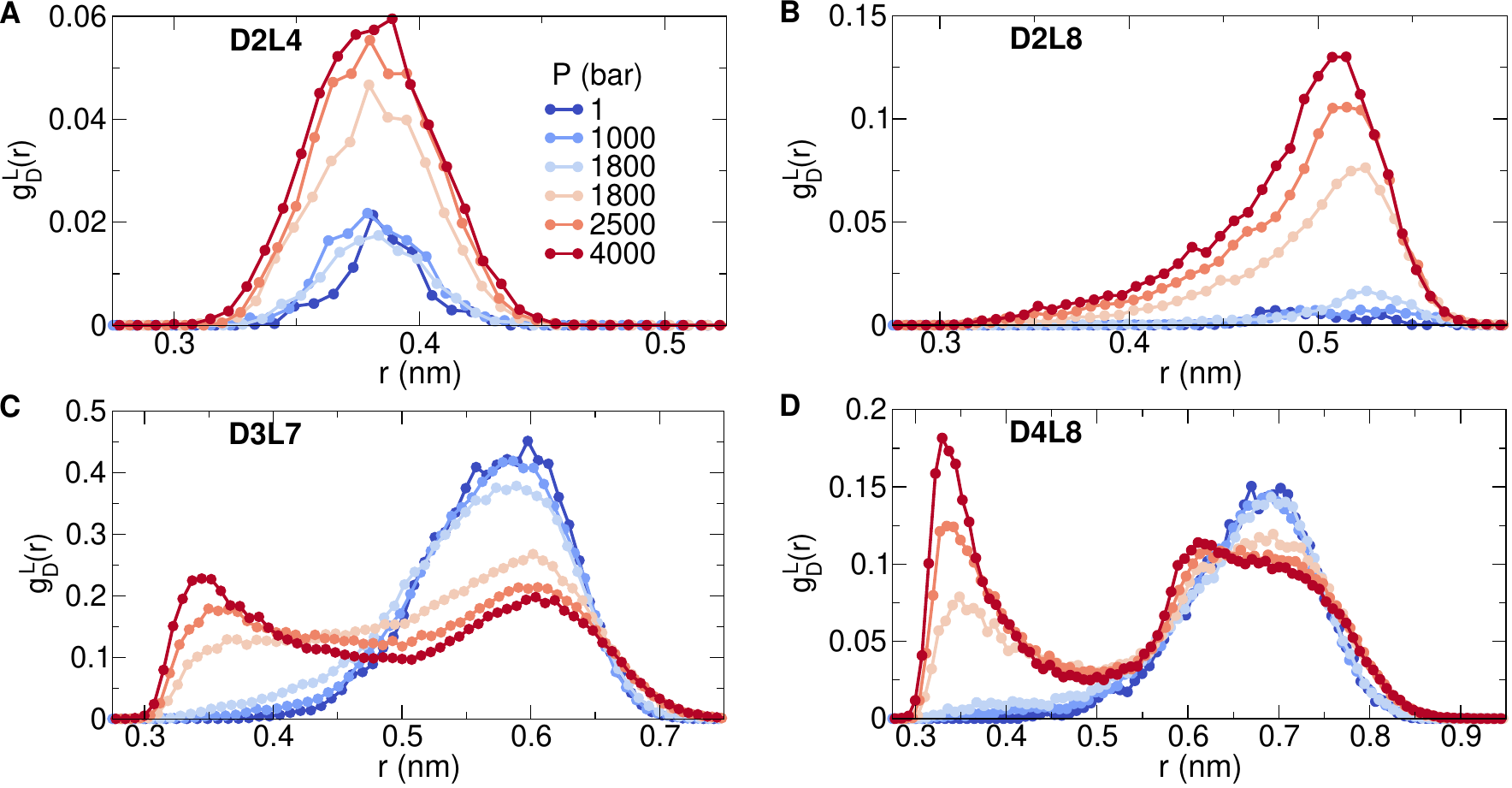}
    \caption{Closeup on the pressure dependence of the $g(r)$ contributions from (\textbf A) D2L4, (\textbf B) D2L8, (\textbf C) D3L7 and (\textbf D) D4L8 rings along the \SI{188}{\K} isotherm. Legend shared among all panels.
    Salient characteristics of these distributions are the clear distinctive peaks arising from D3L7 and D4L8 rings in the interstitial region of the HDL; further, D2L8 rings are essentially a unique feature of the high-density phase.}
    \label{fig:RDF-Rings-Closeup}
\end{figure*}

To highlight these important changes in the topology of the HB network,
we show the pressure dependence of the four most interesting contributions in  Fig.~\ref{fig:RDF-Rings-Closeup}  from ambient pressure up to
\SI{4000}{\bar}. The figures clearly indicate the onset of peaks in these distributions in the HDL phase, hence hinting at a potential definition for a local order parameter of the LDL-HDL transition. 
In particular, the signal from D2L8 rings(panel B) is essentially zero in LDL and significantly different from zero in HDL; so is the signal from D3L7 and D4L8 rings(panels C, D) in the regions \( r < \SI{0.45}{\nano\m} \) and \( r < \SI{0.5}{\nano\m} \), respectively, with the latter showing a much more polished peak.
Tetragonal (D2L4) rings(panel A) also show a similar two-state behavior, but they are not completely absent in the LDL. Interestingly, there is no significant pressure-induced change in the shape of their distribution within each phase.
A connection between the presence of tetragonal rings and interstitial molecules was already hypothesized, but not fully developed, by \citet{svishchev1993structure}. However their concentration is quite small, so they provide only a minor contribution to the $g(r)$, comparable to that of network defects(Fig.~\ref{fig:gr}B).

The existence of pairs of molecules lying at intermediate distances between the first and second peaks of the \( g(r) \) as a result of the formation of long folded rings is thus a structural signature of the HDL.

\begin{figure*}[t]
    \centering
    \includegraphics[width=\textwidth]{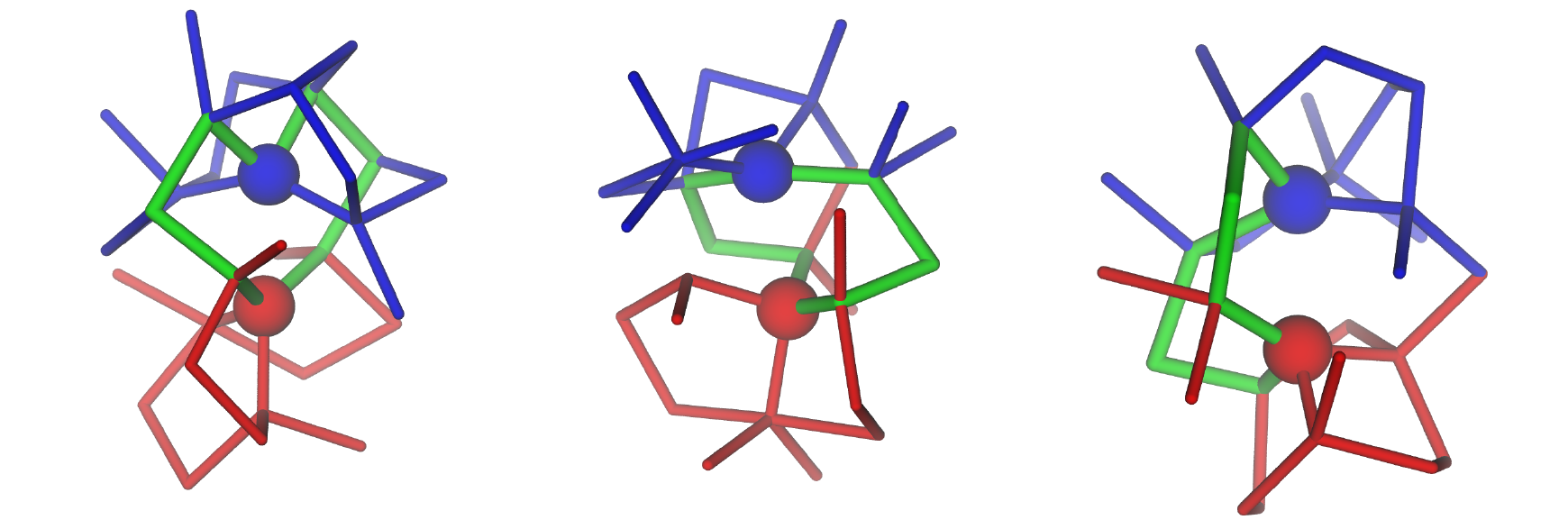}
    \caption{Networks departing from two close-by molecules at chemical distance
    \( D=4 \) in a ring of length \( L=8 \).
    The two reference molecules are identified by blue and red spheres and are at distances \( r < \SI{0.37}{\nano\m} \); each molecule's HB network up to the second generation is represented by blue and red sticks respectively. The ring is identified by green sticks.
    Configurations sampled from simulation data in the HDL at \( P=\SI{1800}{\bar} \).}
    \label{fig:NetworkD4L8}
\end{figure*}

Fig.~\ref{fig:NetworkD4L8} offers a visual representation of the local network structures around close-by($r<\SI{0.37}{\nano\m}$) pairs of molecules at \( D = 4 \) in 8-membered rings, which are arguably the most significant feature of the HDL.
The two reference molecules can be considered as the seeds of two tetrahedral networks which retain their independence up to the second generation. These configurations show different aspects of the high density structures. In the left panel, the ring has a re-entrant cusp-like apex so that the two molecules can get close to each other by interpenetrating their networks (although only on a short length scale); while clearly limited in space by disorder, this type of interpenetration might be reminiscent of the ``full" network interpenetration observed in the crystalline phases of ice VI and VII. In the middle panel the ring displays a folded and strongly non-planar structure, so that molecules on opposite sides of the ring end up close together; in this case, the networks appear to grow in opposite directions, as if repelling each other.
A similar repulsive effect is seen, perhaps even more vividly, in the right panel;
this network repulsion could be suggestively compared to the structural distorsions that the HBs undergo when facing a hydrophobic surface.
We also note that the 8-membered ring connecting the two molecules can be degenerate.



\section{Conclusions}

We have provided a quantitative description of the structural
changes taking place at the liquid-liquid transition in the 
TIP4P/Ice model, a model for which the existence of  a genuine second order phase transition, characterized by fluctuations consistent with the Ising universality class in three dimensions has been recently demonstrated~\cite{debenedetti2020second}. 
We have focused on an isotherm below the critical temperature, 
for which, with simulations longer than several \si{\mu\s}, it is possible to
equilibrate  configurations in metastable equilibrium. More specifically, we have focused on the structural and topological changes taking place
at the LL transition, supported by the possibility to contrast two 
\SI{15}{\mu\s} runs with the same temperature and pressure but different density.  

We have  capitalized on the possibility to define unambiguously the
existing HBs, a possibility offered by the low $T$ at which the LL transition takes place, improved by the minimization procedure that eliminates vibrational distortion.   The important result is the observation that on both sides of the transition, essentially all \ce{H} atoms are involved in HBs.  Again, this is a result enforced by the low \( T \), which makes the loss of an HB (with its characteristic energy  significantly larger than the thermal energy) a rare event.
The large majority of the possible HBs are distributed in the characteristic two-donors two-acceptors tetrahedral state, with a minority of (equally numerous) three and five coordinated particles. The formation of these network defect pairs preserves the total number of HB and it has thus a relatively minor energetic cost.
At coexistence, both low and high density water can be described as a network of (mostly) four-coordinated molecules. 

Finally, we have analysed the changes in the \ce{H}-bond network across the transition with the aim of better understanding the structural origin of the transition itself.  Indeed, there is consensus (supported by experimental neutron and X-rays measurements in the high- and low-density amorphous structures)~\cite{amann2016colloquium}  that the dominant structural change between low- and high-density amorphous ice is the presence at high density of interstitial molecules located around \( r \approx \SI{0.35}{\nano\m} \),
in between the first and second shell of the structure expected for a random tetrahedral network. These interstitial molecules have been thought to originate from the collapse of the second shell. 
By partitioning the radial distribution function in  contributions associated to different chemical distances (defined as the number of HBs separating two molecules), and at an even more sophisticated level separating each chemical distance in contributions arising from rings of different length, we have been able to show that the ``typical" interstitial molecule is tetrahedrally coordinated and is connected to
the central molecule by chemical distances rarely of two (these ones associated to the formation of rings of length four) but mostly three and four.
Thus, rings of length seven and longer adopt folded and elongated shapes to bring two molecules on opposite ring-sides close-by in space.   
The presence of these folded long rings constitutes a characteristic property of the high-density liquid phase.

Pairs of molecules at the interstitial distance \( r \approx \SI{0.35}{\nano\m} \) 
connected by these long rings have both distinct and well-defined first and quite often also second bond-coordination shells. 
In other words, the two close-by molecules facing each other can be considered the 
starting points of two tetrahedral networks, which retain their independence
up to the second generation and which grow on opposite sides, as if repelling  each other, as vividly shown in Fig.~\ref{fig:NetworkD4L8}.

The results presented in this manuscript demonstrate that the network topology contains the information for deciphering the structural differences between the low- and high-density liquid, and their significance also naturally extends to the description of low- and high-density amorphous forms of water, these glasses being --- in the presence of a LL critical point --- the LDL and HDL arrested counterparts. The results shown in Fig.~\ref{fig:RDF-Rings-Closeup} make clear that well defined peaks in the radial distribution function of the HDL phase can be discerned once the contributions are separated on the basis of ring lengths and chemical distances. This offers the possibility to identify local structures which are unambiguously part of the HDL state. 
Future work may indicate if this classification can be extended to supercritical conditions up to ambient temperature and pressure.

Finally we notice that particle-rings of length eight (defined as described in sec.~\ref{sec:particlering}) exist only in the HDL phase. An order parameter to investigate the nucleation of one liquid phase into the other one could possibly be designed on the basis of the present observation.

{\bf Acknowledgements}
\noindent
JR  acknowledges support from the European Research Council Grant DLV-759187.  FS and RF acknowledge support from  MIUR-PRIN (Grant No. 2017Z55KCW). 

{\bf Data availability}
\noindent
The data that support the findings of this study are available within the article.  
Additional data (GROMACS input files) are available from the corresponding author upon reasonable request.


%

\end{document}